\documentclass[twocolumn,american]{revtex4-2}
\usepackage[T1]{fontenc}
\usepackage[utf8]{luainputenc}
\usepackage{amsmath}
\usepackage{amssymb}
\usepackage{babel}
\begin{document}
\title{Kondo Impurities at a Finite Concentration of Impurities}
\author{Garry Goldstein}
\address{garrygoldsteinwinnipeg@gmail.com}
\begin{abstract}
In this work we study the Kondo impurity problem - at a finite concentration
of impurities. We identify two parameter regimes for the Kondo impurity
problem. 1) The single impurity limit, where the concentration of
Kondo impurities is so low that the background scattering mechanisms
(non-magnetic impurities, Umklapp scattering, etc.) of the metal considered
are the dominant conduction electron scattering mechanisms at zero
temperature. 2) The dilute impurity system limit where the concentration
of magnetic impurities is such that they form the dominant mechanism
of conduction electron scattering at zero temperature of the metal
in question (this is accompanied by a variety of easily detectable
Kondo signatures (resistance minimum, specific heat measurements,
magnetization as a function of external magnetic field, conduction
electron dephasing rates as well as ARPES, RIXS and NMR spectroscopies))
while still being very dilute. Most theoretical efforts are currently
in regime where a single isolated impurity is considered - regime
1) while most experimental efforts are in regime 2). We present analytical
evidence that this explains the well known discrepancy between experiment
and theory as to the value of the Kondo temperature. We find that
the ratio between the two Kondo temperatures in regime 1) and regime
2) is given by: $\mathcal{R}=\exp\left[\frac{\pi^{2}\rho v_{F}}{2k_{F}^{2}Vol}\right]$
where $\rho$ is the density of states, $v_{F}$ is the fermi velocity,
and $k_{F}$ is the Fermi wavevector and $Vol$ is the volume of a
unit cell. We note that there is no dependence on the impurity concentration
in this ratio so it is possible to define a single Kondo temperature
for limit 2) for the dilute Kondo impurity system. In this work we
present results within the Reed-Newns Kondo meanfield approximation
and to leading order of the linked cluster expansion.
\end{abstract}
\maketitle

\section{Introduction}\label{sec:Intorduction}

The Kondo impurity problem has a long and rich history. The immense
modern interest in the Kondo problem was first sparked by the experimental
observation that metals, with small concentrations of magnetic impurities,
have resistance minimums \citep{White_1979}. The resistance minimum
is the temperature (which turns to be related to but different from
the Kondo temperature) where the resistance of the metal to DC electrical
current is minimized with respect to temperature. Kondo was able to
explain this phenomenon through a perturbation theory calculation
of spin flip scattering of conduction electrons by a single impurity
spin (Kondo impurity) \citep{Bruus_2004,Hewson_1993,Kondo_1964}.
He found that the the resistance increases logarithmically with temperature
due to this electron spin scattering mechanism and for low temperatures
leads to the resistance minimum. These calculations were further confirmed
when a ``poor man's scaling'' calculation for the single Kondo impurity,
which was done by Phillip Anderson, showing a divergence (as a function
of energy under renormalization group flow) of the effective antiferromagnetic
Heisenberg interaction $J$ between the impurity spins and the conduction
electron spins \citep{Coleman_2015,Hewson_1993,Anderson_1969,Anderson_1970,Anderson_1970-1,Anderson_1971,Anderson_1973}.
This lead to an effective Kondo temperature defined at the energy
scale of this divergence. There is no order parameter for the Kondo
problem (the spin impurity and the conduction electrons form a spin
singlet - which does not break any symmetries) so this is a crossover
not a phase transition. This calculation was further supplemented
by numerical renormalization group calculations of the Kondo impurity
model, done by Kenneth Wilson, which showed similar results \citep{Hewson_1993,Wilson_1975}.
Furthermore Reed and Newns \citep{Anderson_1981,Reed_1983,Coleman_1983}
introduced a path integral formulation of the Kondo model leading
to a meanfield solution of the Kondo impurity and the Kondo lattice
models. It was shown that Kondo crossover temperature given by \citep{Coleman_2015,Hewson_1993}:
\begin{equation}
T_{K}^{L}=D\exp\left(\frac{-1}{N\rho J}\right)\label{eq:Kondo_temperature-1}
\end{equation}
for both the impurity and the lattice (analytically they are believed
to be the same temperature). Despite this, here we use $L$ to denote
the lattice (or the single impurity limit (see below)) not to conflict
with our expression for the dilute impurity system Kondo temperature
given in Eq. (\ref{eq:Kondo_temperature-2}). Here $J$ is the Kondo
coupling, $\rho$ is the density (per spin species) of states at the
Fermi energy, $D$ is the bandwidth and $N$ is the spin degeneracy.

One of the surprising experimental results of the Kondo problem is
that the Kondo temperature (the temperature where magnetic impurities
form spin singlets with the conduction electrons) is markedly different
between the Kondo lattice and the dilute Kondo impurity system despite
all theoretical calculations indicating to the contrary - which show
that the isolated Kondo impurity (studied theoretically) and the Kondo
lattice have the same Kondo temperatures \citep{Coleman_2015}. However,
experimentally the Kondo temperature for the Kondo lattice is about
one order of magnitude greater then that of the dilute impurity system.
Typical values are 50K and 5K respectively for the lattice and dilute
impurity system Kondo temperatures \citep{Lonzarich_2016,Yang_2008}
(there is a lot of variability for these numbers \citep{Lonzarich_2016,Yang_2008}).
We again emphasize directly contradicts current quantitative theoretical
considerations (Eq. (\ref{eq:Kondo_temperature-1})). Indeed the authors
of \citep{Lonzarich_2016} qualitatively postulated that the lattice
has a collective coherence effect where the hybridization (coherence)
between the Kondo impurities and conduction electrons (on a lattice)
is somehow collectively enhanced by the presence of a regular array
of these impurities - a new type of collective phenomena effect. Here,
instead, we consider the possibility that the presence of a dilute
concentration of magnetic impurities lowers the Kondo temperature
of the system below what would be expected by single impurity arguments
alone. One can, of course, argue that one can consider a very dilute
system of impurities where each individual magnetic impurity is very
well isolated from the rest of the magnetic impurities \citep{Coleman_2015,Hewson_1993}.
With this argument the effect proposed here seems impossible and it
seems that it is possible to work explicitly in the single impurity
limit. However this very dilute limit is not what practically happens
experimentally \citep{Hewson_1993,Lonzarich_2016}. Indeed for practical
studies of magnetic impurities the impurity concentration is high
enough that the Kondo mechanism is the dominant scattering mechanism
of conduction electrons at zero temperature. Indeed - due to presence
of the resistance minimum \citep{Hewson_1993}, the conduction electron
lifetimes as extrapolated from NMR signals and neutron scattering
in metals with magnetic impurities \citep{Hewson_1993} as well as
dephasing times in mesoscopic wires with magnetic impurities \citep{Gougham_2000,Mallet_2006,Pierre_2003,Saminadayar_2007,Schopfter_2003,Micklitz_2006}
- we know the Kondo impurities can be considered the main source of
decoherence of conduction electrons in a metal for most modern experiments
on Kondo impurities. We also note that NMR, RIXS, ARPES etc. experiments
require high concentration of magnetic impurities for clear signal
\citep{Hewson_1993}. Alternatively, Kondo impurities at zero temperature
are at the unitarity limit - theoretically - (see Eq. (\ref{eq:Lifetime})
and \citep{Hewson_1993}) and a very small concentration of such -
unitarity limit - impurities becomes the dominant scattering mechanism
in clean metals where the low temperature prevents phonon and Fermi
liquid contributions to the scattering lifetime, similar results are
well known for mesoscopic wires and their dephasing times \citep{Micklitz_2006}.
Here we show that when Kondo impurities are the main source of decoherence
of conduction electrons (there is a resistance minimum and other easily
identifiable Kondo features) the effect of other impurities on a single
impurity does not depend on the impurity concentration (to leading
order in the limit of small impurity concentration) so can be quite
large even for very dilute impurities (of concentration above the
main scattering mechanism requirement cutoff). As such we introduce
two limits of the Kondo impurity problem 1) the single impurity limit
(where there is no easily visible experimental signature of the Kondo
effect) however Kondo impurities are not the main source of conduction
electron decoherence at zero temperature and 2) the dilute impurity
system limit (where there is a resistance minimum and other signatures
of Kondo physics) and the magnetic impurities are the dominant source
of decoherence of conduction electrons at zero temperature. Most experiments
on magnetic impurities are in the latter limit while most theory is
in the former limit - explaining various discrepancies, in particular
in the value of the Kondo temperature.

Before proceeding with detailed calculations we would like to qualitatively
argue about why these two limits (regimes 1) and 2)) are markedly
different. We first note that the Kondo effect is about coherence,
indeed it is about the situation where the impurity electron forms
a spin singlet with the conduction electron at the site of the impurity;
so qualitatively Kondo impurities would have to interact coherently
to interact at all. While low concentration of impurities leads to
weak interactions between nearest impurity neighbor impurities, each
impurity interacts with a large number of other impurities (which
in this work we study analytically to leading order within the linked
cluster expansion \citep{Pathria_2011}). The range where this interaction
is effective is inversely proportional to the decoherence rate of
a conduction electron traveling through the metal and therefore inversely
proportional to the concentration of Kondo impurities (in limit 2)
but not limit 1)). Direct detailed calculation within the Reed-Newns
Kondo meanfield approximation and to leading order within the linked
cluster approximation, see Sections \ref{sec:Main-Setup}, \ref{sec:Main-Calculations}
and \ref{sec:Impurity-Averaging}, show the two effects cancel exactly.
That is, the interaction strength between impurities decreases exactly
at the rate where the number of relevant impurities increases as we
change the impurity concentration leading to the cancellation of impurity
density coefficient in the change of the Kondo temperature relation
(provided the Kondo impurities are the main source of decoherence
of conduction electrons - regime 2)). This leads to a dilute impurity
system Kondo temperature (different then the lattice Kondo temperature
which is the same as the single impurity Kondo temperature) independent
of the concentration of Kondo impurities in the low concentration
limit (but not the single impurity limit - we focus on regime 2)).
We note that this interaction between Kondo impurities is not of Rudernman-Kittel-Kasuya-Yosida
(RKKY) type as its effect does not decrease with large spin degeneracy
$N$ and is mediated by a single particle interaction (we compute
single particle Green's functions rather then two particle ones) \citep{Coleman_2015,Matho_1973,Barzykin_2000,Yosida_1957,Loinaz_1993}.
Furthermore it has a $\sim1/R$ dependence on distance between impurities,
$R$, rather then $1/R^{3}$ as in RKKY \citep{Coleman_2015,Yosida_1957}.
Furthermore it is not an interference effect of the Kondo screening
clouds as the Kondo length $l_{K}\sim v_{F}/T_{K}$ (here $v_{F}$
is the fermi velocity) is not relevant to the physics of this interaction
\citep{Najati_2017,Borzenets_2020,Sorensen_1996}. The author knows
no clear analogies between this interaction and those found in the
literature. The Kondo temperature of the dilute magnetic impurity
system (regime 2)) is then given by: 
\begin{equation}
T_{K}^{I}=D\exp\left[-\frac{1}{N}\left(\frac{1}{\rho J}+N\frac{\pi^{2}\rho v_{F}}{2k_{F}^{2}Vol}\right)\right]\label{eq:Kondo_temperature-2}
\end{equation}
Where $I$ stands for impurity. Where $k_{F}$ is the Fermi wavevector
and $Vol$ is the volume of a unit cell.This leads to a decrease of
the Kondo temperature of the dilute impurity system relative to that
of the Kondo lattice. The ratio of the two temperatures is given by:
\begin{equation}
\mathcal{R}=\frac{T_{K}^{L}}{T_{K}^{I}}=\exp\left[\hbar\frac{\pi^{2}\rho v_{F}}{2k_{F}^{2}Vol}\right]\label{eq:Ratio}
\end{equation}
Putting in experimentally relevant values of $v_{F}=10^{6}m/s$, $k_{F}=2\pi\times0.2\cdot10^{10}m^{-1}$,
$Vol=3\cdot10^{-29}m^{3}$ and $\rho=1eV^{-1}$ we obtain: $\mathcal{R}\sim2-3$
which is a little too small but perhaps the right order of magnitude.
Furthermore there is a great variability in this ratio $\mathcal{R}$
both experimentally \citep{Lonzarich_2016} and now theoretically
(indeed our estimates for $\rho,\,k_{F},\,v_{F},\,Vol$ are crude
and changes in these parameters appear in the exponent). Alternatively
we note that writing $\rho\cdot\left(\hbar v_{F}k_{F}\right)=\rho\cdot D=\frac{1}{2}$
(for the uniform density limit) and $\frac{4\pi}{3}k_{F}^{3}=\frac{1}{2}\frac{\left(2\pi\right)^{3}}{Vol}$
(the Fermi surface is spherically symmetric and the metal is at half
filling) we obtain $\mathcal{R}\sim\exp\left(\frac{1}{12}\right)$
which is too small and too universal. However given the crudeness
of our approximations we feel it is acceptable (this is a pioneering
study of this effect as we find no analogies of it in the literature).
We now proceed to calculate this in detail.

\section{Main Setup}\label{sec:Main-Setup}

In this work we consider the dilute Kondo impurity system with quenched
impurities. The system consists of conduction electrons and spin impurities
at random locations. The Hamiltonian for the system is given by: 
\begin{equation}
H_{K}=\sum_{\sigma}\epsilon\left(\mathbf{k}\right)c_{\sigma}^{\dagger}\left(\mathbf{k}\right)c_{\sigma}\left(\mathbf{k}\right)+J\sum_{\mathbf{r}_{i}}\vec{S}_{i}\cdot\vec{s}_{\mathbf{r}_{i}}\label{eq:Kondo-1}
\end{equation}
Here $\mathbf{r}_{i}$ are the random positions of the impurities,
$S_{i}$ are the localized spin impurities and $s_{\mathbf{r}_{i}}$
are the spin operators for the conduction electrons at the impurity
locations $\mathbf{r}_{i}$ and $J$ is the antiferromagnetic Kondo
coupling. We will assume a uniform density of states $\rho$ and a
bandwidth $D$. We now follow \citep{Coleman_2015} (section 17.5)
and perform a large $N$ Reed-Newns path integral and obtain the meanfield
expression for the Helmhotlz free energy of the system. We note that
because the Kondo effect is about the formation of spin singlets (which
have no spatial directions) we may use the same quantization axis
for all the impurities without loss of generality. We obtain that
\citep{Coleman_2015}: 
\begin{equation}
F=\sum_{i}\left[\frac{\left|V_{i}\right|^{2}}{J}-Q\lambda_{i}\right]-Nk_{B}T\ln\left[det\left[G\left(\omega_{n},\left\{ \mathbf{r}_{i}\right\} \right)\right]\right]\label{eq:Helmhotlz}
\end{equation}
Here $V_{i}$ are the hybridizations of the impurities at $\mathbf{r}_{i}$,
$\lambda_{i}$ are the Lagrange multipliers used to enforce impurity
occupation numbers $q=\frac{Q}{N}$ (we will work at $q=\frac{1}{2}$).
Furthermore $k_{B}$ is the Boltzmann constant, $\omega_{n}$ are
Matsubara frequencies and $T$ is the temperature. We will eventually
take the zero temperature limit. 
\begin{widetext}
Where we have that $G\left(\omega_{n},\left\{ \mathbf{r}_{i}\right\} \right)=$
\begin{equation}
\left(\begin{array}{cccccccc}
\epsilon\left(\mathbf{k}_{1}\right)-i\omega_{n} & 0 & \cdots & 0 & V_{1}^{*}\exp\left(-i\mathbf{k}_{1}\cdot\mathbf{r}_{1}\right) & V_{2}^{*}\exp\left(-i\mathbf{k}_{1}\cdot\mathbf{r}_{1}\right) & \cdots & V_{\mathcal{N}}^{*}\exp\left(-i\mathbf{k}_{1}\cdot\mathbf{r}_{\mathcal{N}}\right)\\
0 & \epsilon\left(\mathbf{k}_{2}\right)-i\omega_{n} & \ddots & \vdots & V_{1}^{*}\exp\left(-i\mathbf{k}_{2}\cdot\mathbf{r}_{1}\right) & V_{2}^{*}\exp\left(-i\mathbf{k}_{2}\cdot\mathbf{r}_{1}\right) & \cdots & V_{\mathcal{N}}^{*}\exp\left(-i\mathbf{k}_{2}\cdot\mathbf{r}_{\mathcal{N}}\right)\\
\vdots & \ddots & \ddots & 0 & \vdots & \vdots & \ddots & \vdots\\
0 & \cdots & 0 & \epsilon\left(\mathbf{k}_{t}\right)-i\omega_{n} & V_{1}^{*}\exp\left(-i\mathbf{k}_{t}\cdot\mathbf{r}_{1}\right) & V_{2}^{*}\exp\left(-i\mathbf{k}_{t}\cdot\mathbf{r}_{1}\right) & \cdots & V_{\mathcal{N}}^{*}\exp\left(-i\mathbf{k}_{t}\cdot\mathbf{r}_{\mathcal{N}}\right)\\
V_{1}\exp\left(i\mathbf{k}_{1}\cdot\mathbf{r}_{1}\right) & V_{1}\exp\left(i\mathbf{k}_{2}\cdot\mathbf{r}_{1}\right) & \cdots & V_{1}\exp\left(i\mathbf{k}_{t}\cdot\mathbf{r}_{1}\right) & \lambda_{1}-i\omega_{n} & 0 & \cdots & 0\\
V_{2}\exp\left(i\mathbf{k}_{1}\cdot\mathbf{r}_{1}\right) & V_{2}\exp\left(i\mathbf{k}_{2}\cdot\mathbf{r}_{1}\right) & \cdots & V_{2}\exp\left(i\mathbf{k}_{t}\cdot\mathbf{r}_{1}\right) & 0 & \lambda_{2}-i\omega_{n} & \ddots & \vdots\\
\vdots & \vdots & \ddots & \vdots & \vdots & \ddots & \ddots & 0\\
V_{\mathcal{N}}\exp\left(i\mathbf{k}_{1}\cdot\mathbf{r}_{\mathcal{N}}\right) & V_{\mathcal{N}}\exp\left(i\mathbf{k}_{2}\cdot\mathbf{r}_{\mathcal{N}}\right) & \cdots & V_{\mathcal{N}}\exp\left(i\mathbf{k}_{t}\cdot\mathbf{r}_{\mathcal{N}}\right) & 0 & \cdots & 0 & \lambda_{\mathcal{N}}-i\omega_{n}
\end{array}\right)\label{eq:Matrix}
\end{equation}
Here $t$ is the total number of $\mathbf{k}$ values considered and
we assume $\mathcal{N}$ spins. We now do a linked cluster expansion
for the value of $\ln\left[det\left[G\left(\omega_{n},\left\{ \mathbf{r}_{i}\right\} \right)\right]\right]$
\citep{Pathria_2011}. That is we define 
\begin{equation}
\Delta^{0}\ln\left[det\left[G\left(\omega_{n},\left\{ \mathbf{r}_{i}\right\} \right)\right]\right]=\ln\left[det\left[G^{0}\left(\omega_{n},\left\{ \mathbf{r}_{i}\right\} \right)\right]\right]\label{eq:Determinant}
\end{equation}
Where: 
\begin{equation}
G^{0}\left(\omega_{n},\left\{ \mathbf{r}_{i}\right\} \right)=\left(\begin{array}{ccccc}
\epsilon\left(\mathbf{k}_{1}\right)-i\omega_{n} & 0 & 0 & \cdots & 0\\
0 & \epsilon\left(\mathbf{k}_{2}\right)-i\omega_{n} & \cdots & \cdots & \vdots\\
\vdots & 0 & \ddots & \ddots & \vdots\\
0 & \ddots & \ddots & \ddots & 0\\
0 & 0 & \cdots & 0 & \epsilon\left(\mathbf{k}_{t}\right)-i\omega_{n}
\end{array}\right)\label{eq:Main_matrix}
\end{equation}
Now we define: 
\begin{equation}
\Delta^{i}\ln\left[det\left[G\left(\omega_{n},\left\{ \mathbf{r}_{i}\right\} \right)\right]\right]=\ln\left[det\left[G^{i}\left(\omega_{n},\left\{ \mathbf{r}_{i}\right\} \right)\right]\right]-\Delta^{0}\ln\left[det\left[G\left(\omega_{n},\left\{ \mathbf{r}_{i}\right\} \right)\right]\right]\label{eq:Delta_i}
\end{equation}
Where: 
\begin{equation}
G^{i}\left(\omega_{n},\left\{ \mathbf{r}_{i}\right\} \right)=\left(\begin{array}{cccccc}
\epsilon\left(\mathbf{k}_{1}\right)-i\omega_{n} & 0 & 0 & \cdots & 0 & V_{i}^{*}\exp\left(-i\mathbf{k}_{1}\cdot\mathbf{r}_{1}\right)\\
0 & \epsilon\left(\mathbf{k}_{2}\right)-i\omega_{n} & \cdots & \cdots & \vdots & V_{i}^{*}\exp\left(-i\mathbf{k}_{2}\cdot\mathbf{r}_{1}\right)\\
\vdots & 0 & \ddots & \ddots & \vdots & \vdots\\
0 & \ddots & \ddots & \ddots & 0 & \vdots\\
0 & 0 & \cdots & 0 & \epsilon\left(\mathbf{k}_{t}\right)-i\omega_{n} & V_{i}^{*}\exp\left(-i\mathbf{k}_{t}\cdot\mathbf{r}_{1}\right)\\
V_{i}\exp\left(i\mathbf{k}_{1}\cdot\mathbf{r}_{1}\right) & V_{i}\exp\left(i\mathbf{k}_{2}\cdot\mathbf{r}_{1}\right) & \cdots & \cdots & V_{i}\exp\left(i\mathbf{k}_{t}\cdot\mathbf{r}_{1}\right) & \lambda_{i}-i\omega_{n}
\end{array}\right)\label{eq:Matrix_i}
\end{equation}
We then define: 
\begin{equation}
\Delta^{ij}\ln\left[det\left[G\left(\omega_{n},\left\{ \mathbf{r}_{i}\right\} \right)\right]\right]=\ln\left[det\left[G^{ij}\left(\omega_{n},\left\{ \mathbf{r}_{i}\right\} \right)\right]\right]-\Delta^{i}\ln\left[det\left[G\left(\omega_{n},\left\{ \mathbf{r}_{i}\right\} \right)\right]\right]-\Delta^{j}\ln\left[det\left[G\left(\omega_{n},\left\{ \mathbf{r}_{i}\right\} \right)\right]\right]-\Delta^{0}\ln\left[det\left[G\left(\omega_{n},\left\{ \mathbf{r}_{i}\right\} \right)\right]\right]\label{eq:Main_step}
\end{equation}
Where 
\begin{equation}
G^{ij}\left(\omega_{n},\left\{ \mathbf{r}_{i}\right\} \right)=\left(\begin{array}{ccccccc}
\epsilon\left(\mathbf{k}_{1}\right)-i\omega_{n} & 0 & 0 & \cdots & 0 & V_{i}^{*}\exp\left(-i\mathbf{k}_{1}\cdot\mathbf{r}_{1}\right) & V_{j}^{*}\exp\left(-i\mathbf{k}_{1}\cdot\mathbf{r}_{1}\right)\\
0 & \epsilon\left(\mathbf{k}_{2}\right)-i\omega_{n} & \cdots & \cdots & \vdots & V_{i}^{*}\exp\left(-i\mathbf{k}_{2}\cdot\mathbf{r}_{1}\right) & V_{j}^{*}\exp\left(-i\mathbf{k}_{2}\cdot\mathbf{r}_{1}\right)\\
\vdots & 0 & \ddots & \ddots & \vdots & \vdots & \vdots\\
0 & \ddots & \ddots & \ddots & 0 & \vdots & \vdots\\
0 & 0 & \cdots & 0 & \epsilon\left(\mathbf{k}_{t}\right)-i\omega_{n} & V_{i}^{*}\exp\left(-i\mathbf{k}_{t}\cdot\mathbf{r}_{1}\right) & V_{j}^{*}\exp\left(-i\mathbf{k}_{t}\cdot\mathbf{r}_{1}\right)\\
V_{i}\exp\left(i\mathbf{k}_{1}\cdot\mathbf{r}_{1}\right) & V_{i}\exp\left(i\mathbf{k}_{2}\cdot\mathbf{r}_{1}\right) & \cdots & \cdots & V_{i}\exp\left(i\mathbf{k}_{t}\cdot\mathbf{r}_{1}\right) & \lambda_{i}-i\omega_{n} & 0\\
V_{j}\exp\left(i\mathbf{k}_{1}\cdot\mathbf{r}_{1}\right) & V_{j}\exp\left(i\mathbf{k}_{2}\cdot\mathbf{r}_{1}\right) & \cdots & \cdots & V_{j}\exp\left(i\mathbf{k}_{t}\cdot\mathbf{r}_{1}\right) & 0 & \lambda_{j}-i\omega_{n}
\end{array}\right)\label{eq:Matrix_ij}
\end{equation}
Then we have that \citep{Pathria_2011}:
\begin{equation}
\ln\left[det\left[G\left(\omega_{n},\left\{ \mathbf{r}_{i}\right\} \right)\right]\right]=\Delta^{0}\ln\left[det\left[G\left(\omega_{n},\left\{ \mathbf{r}_{i}\right\} \right)\right]\right]+\sum_{i}\Delta^{i}\ln\left[det\left[G\left(\omega_{n},\left\{ \mathbf{r}_{i}\right\} \right)\right]\right]+\sum_{i<j}\Delta^{ij}\ln\left[det\left[G\left(\omega_{n},\left\{ \mathbf{r}_{i}\right\} \right)\right]\right]+.....\label{eq:Leading_order}
\end{equation}
\end{widetext}

We will not be interested in higher order terms in the linked cluster
expansion in this work \citep{Pathria_2011}.

\section{ Calculation of $\ln\left[det\left[G^{ij}\left(\omega_{n},\left\{ \mathbf{r}_{i}\right\} \right)\right]\right]$}\label{sec:Main-Calculations}

We note that $\ln\left[det\left[G^{0}\left(\omega_{n},\left\{ \mathbf{r}_{i}\right\} \right)\right]\right]$
is an overall energy shift that will cancel everywhere. The calculation
of $\ln\left[det\left[G^{i}\left(\omega_{n},\left\{ \mathbf{r}_{i}\right\} \right)\right]\right]$
is well known with the final result that \citep{Coleman_2015}: 
\begin{align}
 & \ln\left[det\left[G^{i}\left(\omega_{n},\left\{ \mathbf{r}_{i}\right\} \right)\right]\right]\nonumber \\
 & =\ln\left[det\left[G^{0}\left(\omega_{n},\left\{ \mathbf{r}_{i}\right\} \right)\right]\right]\nonumber \\
 & +\frac{1}{\pi k_{B}T}Im\left[\int_{-D}^{D}d\omega f\left(\omega\right)\ln\left(-\omega+\lambda_{i}+i\Delta_{i}\right)\right]\nonumber \\
 & =\ln\left[det\left[G^{0}\left(\omega_{n},\left\{ \mathbf{r}_{i}\right\} \right)\right]\right]+\frac{1}{\pi k_{B}T}Im\left[\xi_{i}\ln\left(\frac{\xi_{i}}{eD}\right)\right]\label{eq:Pierce}
\end{align}
Where 
\begin{equation}
\Delta_{i}=\pi\left|V_{i}\right|^{2}\rho,\:\xi_{i}=\lambda_{i}+i\Delta_{i}\label{eq:definitions}
\end{equation}
Here $f\left(\omega\right)$ is Fermi-Dirac distribution, where we
took the zero temperature limit. We now move on to calculating $\ln\left[det\left[G^{ij}\left(\omega_{n},\left\{ \mathbf{r}_{i}\right\} \right)\right]\right]$.
Now we use the relationship that \citep{Coleman_2015}
\begin{equation}
\det\left(\begin{array}{cc}
D & C\\
B & A
\end{array}\right)=\det\left(D\right)\cdot\det\left(A-BD^{-1}C\right)\label{eq:Convenient}
\end{equation}
to obtain \citep{Coleman_2015}: 
\begin{align}
 & \ln\left[det\left[G^{ij}\left(\omega_{n},\left\{ \mathbf{r}_{i}\right\} \right)\right]\right]\nonumber \\
 & =\ln\left[det\left[G^{0}\left(\omega_{n},\left\{ \mathbf{r}_{i}\right\} \right)\right]\right]+\ln\left[det\left[M^{ij}\left(\omega_{n},\left\{ \mathbf{r}_{i}\right\} \right)\right]\right]\label{eq:Decomposition}
\end{align}
Where: 
\begin{align}
 & M^{i,j}\left(\omega_{n},\left\{ \mathbf{r}_{i}\right\} \right)\nonumber \\
 & =\left[\begin{array}{cc}
i\omega_{n}+\lambda_{i}+\sum_{\mathbf{k}}\frac{\left|V_{i}\right|^{2}}{i\omega_{n}-\epsilon_{\mathbf{k}}} & \sum_{\mathbf{k}}\frac{V_{i}^{*}V_{j}\exp\left(i\mathbf{k}\cdot\left(\mathbf{r}_{j}-\mathbf{r}_{i}\right)\right)}{i\omega_{n}-\epsilon_{\mathbf{k}}}\\
\sum_{\mathbf{k}}\frac{V_{j}^{*}V_{i}\exp\left(i\mathbf{k}\cdot\left(\mathbf{r}_{i}-\mathbf{r}_{j}\right)\right)}{i\omega_{n}-\epsilon_{\mathbf{k}}} & i\omega_{n}+\lambda_{2}+\sum_{\mathbf{k}}\frac{\left|V_{j}\right|^{2}}{i\omega_{n}-\epsilon_{\mathbf{k}}}
\end{array}\right]\label{eq:M_12}
\end{align}
Now performing the determinant then introducing Matsubara contours
\citep{Coleman_2015} then deforming the contour to the branch cut
across the real axis we obtain \citep{Coleman_2015,Akkermans_2007}:
\begin{align}
 & \ln\left[det\left[M^{i,j}\left(\omega_{n},\left\{ \mathbf{r}_{i}\right\} \right)\right]\right]\nonumber \\
 & =-\frac{1}{\pi k_{B}T}\int_{-D}^{D}d\omega f\left(\omega\right)Im\left[\ln\left[\left(-\omega+\lambda_{i}+i\Delta_{i}\right)\times\right.\right.\nonumber \\
 & \times\left(-\omega+\lambda_{j}+i\Delta_{j}\right)+\nonumber \\
 & \left.\left.+\left|V_{i}\right|^{2}\left|V_{j}\right|^{2}\pi^{2}\rho^{2}\frac{\sin^{2}\left(k_{F}r\right)}{\left(k_{F}r\right)^{2}}\exp\left(-2\frac{\Gamma}{v_{F}}r_{ij}\right)\right]\right]\label{eq:Determinant_final}
\end{align}
Where $r_{ij}=\left|\mathbf{r}_{i}-\mathbf{r}_{j}\right|$. and $\Gamma$
is the decoherence rate of conduction electrons. Where we have re-summed
further by adding a decoherence to the free Green's functions (making
this not strictly a linked cluster expansion), see Eq. (\ref{eq:Calculation}).
Where:
\begin{equation}
\Gamma=\Gamma_{0}+\frac{1}{2\tau_{K}}\label{eq:Kondo_lifetime}
\end{equation}
here $\Gamma_{0}$ is some background decoherence rate due to say
non-magnetic impurities and Umklapp scattering. Where \citep{Hewson_1993}:
\begin{equation}
\tau_{K}^{-1}=\frac{2}{\pi\rho}\frac{\mathcal{N}}{\mathcal{N}_{S}}\label{eq:Lifetime}
\end{equation}
Where $\mathcal{N}_{S}$ is the total number of sites in the crystal
and $\mathcal{N}$ is the total number of impurities. We now introduce
\citep{Coleman_2015} the variables:
\begin{equation}
C_{ij}=\left|V_{i}\right|^{2}\left|V_{j}\right|^{2}\pi^{2}\rho^{2}\frac{\sin^{2}\left(k_{F}r\right)}{\left(k_{F}r\right)^{2}}\exp\left(-2\frac{\Gamma}{v_{F}}r_{ij}\right)\label{eq:Variables}
\end{equation}
Now we specialize to $\xi_{1}=\xi_{2}$, where each impurity interacts
with many impurities and therefore has essentially the same hybridization
(fluctuations do not matter in the dilute limit). We now obtain \citep{Coleman_2015}:
\begin{align}
 & \ln\left[det\left[M^{ij}\left(\omega_{n},\left\{ \mathbf{r}_{i}\right\} \right)\right]\right]\nonumber \\
 & =\frac{1}{\pi k_{B}T}Im\left[\left[\xi+i\sqrt{C_{ij}}\right]\ln\left(\frac{\left[\xi+i\sqrt{C_{ij}}\right]}{eD}\right)+\right.\nonumber \\
 & \left.+\left[\xi-i\sqrt{C_{ij}}\right]\ln\left(\frac{\left[\xi-i\sqrt{C_{ij}}\right]}{eD}\right)\right]\nonumber \\
 & =\Delta^{i}\ln\left[det\left[G\left(\omega_{n},\left\{ \mathbf{r}_{i}\right\} \right)\right]\right]+\Delta^{j}\ln\left[det\left[G\left(\omega_{n},\left\{ \mathbf{r}_{i}\right\} \right)\right]\right]\nonumber \\
 & +\frac{1}{\pi k_{B}T}Im\left[\left(\xi+i\sqrt{C_{ij}}\right)\ln\left(1+i\sqrt{C_{ij}}/\xi\right)\right.\nonumber \\
 & \left.+\left(\xi-i\sqrt{C_{ij}}\right)\ln\left(1-i\sqrt{C_{ij}}/\xi\right)\right]\label{eq:Expansion}
\end{align}
Where we have set $f\left(\omega\right)=\Theta\left(\omega\right)$
the heavy side function and performed the integral in Eq. (\ref{eq:Determinant_final}).
We now perform impurity averaging, since we are interested in the
dilute impurity limit we have that: 
\begin{align}
 & \left(\xi+i\sqrt{C_{ij}}\right)\ln\left(1+i\sqrt{C_{ij}}/\xi\right)+\left(\xi-i\sqrt{C_{ij}}\right)\ln\left(1-i\sqrt{C_{ij}}/\xi\right)\nonumber \\
 & =-\frac{C_{ij}}{\xi}+\frac{1}{6}\frac{C_{ij}^{2}}{\xi^{3}}+...\label{eq:Taylor}
\end{align}
We will stop only at the leading order term $-\frac{C_{ij}}{\xi}$.

\section{Impurity Averaging and Kondo temperature}\label{sec:Impurity-Averaging}

Whereby the Helmholts free energy per impurity is given by \citep{Coleman_2015}:
\begin{widetext}
\begin{align}
F & =\frac{\left|V\right|^{2}}{J}-Q\lambda-\frac{Im}{\pi}\left[N\xi\ln\left(eD/\xi\right)\right]+\frac{N}{2}k_{B}T\left\langle \Delta^{ij}\ln\left[det\left[G\left(\omega_{n},\left\{ \mathbf{r}_{i}\right\} \right)\right]\right]\right\rangle \nonumber \\
 & =\left[\frac{\Delta}{\pi\rho J}-Q\lambda\right]-\frac{Im}{\pi}\left[N\xi\ln\left(eD/\xi\right)\right]-\frac{Im}{\pi}\left[\left|V\right|^{4}\frac{N}{2\xi}\frac{\mathcal{N}}{\mathcal{N}_{S}\cdot Vol}\int_{0}^{\infty}4\pi r^{2}dr\pi^{2}\rho^{2}\frac{\sin^{2}\left(k_{F}r\right)}{\left(k_{F}r\right)^{2}}\exp\left(-2\frac{\Gamma}{v_{F}}r\right)\right]\nonumber \\
 & \cong\left[\frac{\Delta}{\pi\rho J}-Q\lambda\right]-\frac{Im}{\pi}\left[N\xi\ln\left(eD/\xi\right)\right]-Im\left[\frac{N}{2\xi}\frac{\mathcal{N}}{\mathcal{N}_{S}\cdot Vol}\frac{v_{F}}{k_{F}^{2}\left(\Gamma_{0}+\frac{1}{\pi\rho}\frac{\mathcal{N}}{\mathcal{N}_{S}}\right)}\Delta^{2}\right]\label{eq:Impurity}
\end{align}
Here $\left\langle \right\rangle $ denotes spatial averaging. Here
we have assumed that $k_{F}\gg\Gamma$ so that $\sin^{2}\left(k_{F}r\right)$
averages to $\frac{1}{2}$. We note that in many cases $\Gamma_{0}\ll\frac{1}{\pi\rho}\frac{\mathcal{N}}{\mathcal{N}_{S}}$
(we are in regime 2)), furthermore at half filling ($q=\frac{1}{2}$
) we get that $\lambda=0$ \citep{Coleman_2015} and we need minimize:
\begin{align}
F & =\frac{\Delta}{\pi\rho J}-\frac{1}{\pi}N\Delta\ln\left(eD\right)+\frac{1}{\pi}N\Delta\ln\left(\Delta\right)+N\pi\left[\frac{\rho\Delta}{2k_{F}^{2}}\right]\frac{v_{F}}{Vol}\nonumber \\
\frac{dF}{d\Delta} & =\frac{1}{\pi\rho J}-\frac{N}{\pi}\ln\left(D/\Delta\right)+N\pi\frac{\rho}{2k_{F}^{2}}\frac{v_{F}}{Vol}\nonumber \\
0 & =\frac{1}{\rho J}-N\ln\left(D/T_{K}^{I}\right)+N\pi^{2}\frac{\rho}{2k_{F}^{2}}\frac{v_{F}}{Vol}\label{eq:Minimize}
\end{align}
As such we obtain Eq. (\ref{eq:Kondo_temperature-2}).
\end{widetext}

\section{Conclusions}\label{sec:Conclusions}

In this work we studied the Kondo impurity problem at a small, but
finite, concentration of magnetic impurities. We identified two regimes
for the Kondo impurity problem 1) the single impurity regime where
background scattering, as such non-magnetic impurities and Umklapp
scattering, dominate the scattering mechanisms for the conduction
electrons at zero temperature 2) the dilute impurity Fermi liquid
regime where there are many signatures namely a resistance minimum
and magnetic impurities dominate zero temperature scattering. We showed
that most theory is in regime 1) while most experiment is in regime
2). This explains many of the disagreements between theory and experiment
in particular about the value of the Kondo temperature. We took a
step forward by pushing theory into regime 2) which is relevant to
experiments. We showed to leading order in the low impurity concentration
(but in regime 2)) the effects of the impurities on each other do
not depend on impurity concentration (so very low concentrations of
impurities can have profound effects on each other). Indeed the Kondo
effect is about coherence where the impurity spin forms a spin singlet
with the conduction electrons. Therefore, the only way that Kondo
impurities can effect each other is if they coherently interact with
each other. The length where coherent interactions are possible in
determined by the scattering lifetime of conduction electrons at the
Fermi energy or, in the case the impurities are dense enough for the
resistance minimum and other signatures of the Kondo effect to occur,
the concentration of impurities. The two effects: a single impurity
interacts more weakly with its neighbor impurities and there are more
neighbor impurities within a coherence length cancel to leading order
for a dilute impurity system as a function of impurity concentration.
This leads to a Kondo temperature for dilute magnetic impurities which
does not depend on the impurity density and different from the single
magnetic impurity temperature. In the future it would be of interest
to go beyond the leading order in the linked cluster expansion (still
within the Reed-Newns Kondo meanfield) to confirm this result further. 

\appendix

\section{Single Particle Green's functions}\label{sec:Single-Particle-Green's}

Here we would like to derive some properties of single particle Matsubara
free electron Green's functions in real space. The author was unable
to look up the results in a convenient reference and therefore presented
them here. For example the results presented in \citep{Akkermans_2007,Economou_2006}
use free electron dispersion which is unrealistic as it is highly
particle hole asymmetric. Here we correct for this and restore particle
hole symmetry. Consider the single particle Green's functions in real
space given by: 
\begin{align}
G\left(z,\mathbf{r}\right) & =\int\frac{d^{3}\mathbf{k}}{\left(2\pi\right)^{3}}\frac{\exp\left(i\mathbf{k}\cdot\mathbf{r}\right)}{z-\left(\frac{\mathbf{k}^{2}}{2m}-\mu\right)+i\Gamma sgn\left(Im\left(z\right)\right)}\nonumber \\
 & =\int_{0}^{\infty}\int_{0}^{\pi}\int_{0}^{2\pi}\frac{k^{2}dk\sin\left(\theta\right)d\theta d\varphi}{\left(2\pi\right)^{3}}\times\nonumber \\
 & \times\frac{\exp\left(ikr\cos\left(\theta\right)\right)}{z-\left(\frac{k^{2}}{2m}-\mu\right)+i\Gamma sgn\left(Im\left(z\right)\right)}\nonumber \\
 & =\int_{0}^{\infty}\int_{-1}^{1}\frac{k^{2}dkdx}{\left(2\pi\right)^{2}}\frac{\exp\left(ikrx\right)}{z-\left(\frac{k^{2}}{2m}-\mu\right)+i\Gamma sgn\left(Im\left(z\right)\right)}\nonumber \\
 & =2i\int_{0}^{\infty}\frac{kdk}{\left(2\pi\right)^{2}r}\frac{\sin\left(kr\right)}{z-\left(\frac{k^{2}}{2m}-\mu\right)+i\Gamma sgn\left(Im\left(z\right)\right)}\label{eq:No_particle_hole}
\end{align}
Now we restore particle hole symmetry (having done the angular integrals):
\begin{align}
G\left(z,\mathbf{r}\right) & =2im\int_{-D}^{D}\frac{dE}{\left(2\pi\right)^{2}r}\frac{\sin\left(\left[k_{F}+\frac{E}{v_{F}}\right]r\right)}{z-E+i\Gamma sgn\left(Im\left(z\right)\right)}\nonumber \\
 & =i\cdot Im\left[\int_{-D}^{D}\frac{2mdE}{\left(2\pi\right)^{2}r}\frac{\exp\left(i\left[k_{F}+\frac{E}{v_{F}}\right]r\right)}{z-E+i\Gamma sgn\left(Im\left(z\right)\right)}\right]\nonumber \\
 & \cong i\cdot Im\left[\int_{-\infty}^{\infty}\frac{mdE}{2\pi^{2}r}\frac{\exp\left(i\left[k_{F}+\frac{E}{v_{F}}\right]r\right)}{z-E+i\Gamma sgn\left(Im\left(z\right)\right)}\right]\nonumber \\
 & =i\cdot Im\left[\frac{m}{2\pi r}\exp\left(i\left[k_{F}+\frac{z}{v_{F}}\right]r\right)\exp\left(-\frac{\Gamma}{v_{F}}r\right)\right]\nonumber \\
 & \cong i\frac{m}{2\pi r}\sin\left(k_{F}r\right)\exp\left(-\frac{\Gamma}{v_{F}}r\right)sgn\left(Im\left(z\right)\right)\nonumber \\
 & =i\pi\rho\frac{\sin\left(k_{F}r\right)}{k_{F}r}\exp\left(-\frac{\Gamma}{v_{F}}r\right)sgn\left(Im\left(z\right)\right)\label{eq:Calculation}
\end{align}
Where $r=\left|\mathbf{r}\right|$ and
\begin{equation}
\rho=\frac{4\pi k_{F}^{2}}{\left(2\pi\right)^{3}}\cdot\frac{m}{k_{F}}=\frac{k_{F}m}{2\pi^{2}}\label{eq:Density}
\end{equation}
This expression must be used instead of the one typically used for
free green's functions see e.g. \citep{Akkermans_2007,Economou_2006}

\end{document}